\documentclass{epl2}

\usepackage{epsfig}

\title{The initial velocity dependence of the quantum resonance in the delta-kicked rotor}
\author{J-A. Currivan, A. Ullah and M.D. Hoogerland}
\institute{Department of Physics, University of Auckland, Private Bag 92019, Auckland, New Zealand}
\shortauthor{J-A.\,Currivan {\em et. al.}}

\abstract{The influence of an initial momentum on the appearance of ``quantum resonances'' in the delta-kicked rotor system is explored experimentally. We show that for certain initial momenta, a resonance can be negated entirely, whereas at others a resonance can be made to appear. At a larger number of kicks, all resonances are shown to narrow. We show that as a consequence, the individual ``diffraction peaks'' may split. We compare our results to numerical simulations as well as analytical theory. }
\pacs{05.45.Mt}{Quantum chaos; semiclassical methods}
\pacs{32.80.Lg}{Mechanical effects of light on atoms, molecules, and ions}

\begin{document}
%\title{The initial momentum dependence of the quantum resonance in the delta-kicked rotor}
%\author{J-A. Currivan, A. Ullah and M.D. Hoogerland}
%\affiliation{Department of Physics, University of Auckland, Private Bag 92019, Auckland, New Zealand}

%\begin{abstract}
%The influence of an initial momentum on the appearance of ``quantum resonances'' in the delta-kicked rotor system is explored experimentally. We show than for certain initial momenta, a resonance can be negated entirely, whereas at others a resonance can be made to appear. At a larger number of kicks, all resonances are shown to narrow. We compare our results to numerical simulations as well as analytical theory. 
%\end{abstract}
%\pacs{05.45.Mt,32.80.Lg}

\maketitle

%\section{Introduction}
The delta-kicked rotor system has been a paradigm for the study of ``quantum chaos'': quantum systems for which the
classical equivalent exhibits chaos. A model system that has been studied extensively is the ``atom optics'' kicked rotor, which consists of a two-level atom in a detuned, standing wave of laser light. This system has been studied extensively theoretically \cite{izrailev79} and experimentally \cite{raizen99,PhysRevLett.80.4111} for cold atoms originating from a Magneto-Optical Trap (MOT). 
The temperature of the atoms in a MOT is in the order of microKelvins, much larger than the heat input from the recoil of a single photon, and the typical dimensions of a MOT is in the order of a millimetre, much larger than the wavelength of the light in the standing wave. Experiments have demonstrated dynamical localisation \cite{moore95} and the existence of ``quantum resonances'' \cite{oskay00}, depending on the kick period. 

At these quantum resonances, an effect similar to the optical Talbot effect \cite{talbot,deng99} takes place. In this effect, the near field after a light field diffracts from a grating is recreated at a certain distance away from the grating. The analogue in atom optics is observed when a diffracted plane deBroglie wave is diffracted after a certain ``Talbot'' time $T_T=\pi/(2\omega_R)$ the near field is recreated. Here $\omega_R=\frac{\hbar k_L^2}{2m}$ is the recoil frequency, and  we have defined the laser wave number $k_L=2\pi/\lambda$ and the mass of the atom $m$.

When the kick period $T$ corresponds to an integer times the Talbot time ($T=mT_T$, with integer $m$),  
the accumulated phase of the wave function due to moving with a velocity $2\hbar k_L$ is exactly $2\pi$, and the kicks add constructively. This leads to a quadratic growth of the variance of the velocity distribution, or energy, with the number of kicks. For half integer times the Talbot time ($T=(m+\frac{1}{2})T_T$), The kicks add destructively, and the effect of the first kick is canceled by the second. However, for velocity distributions much broader than a photon recoil, an average over the initial velocities leads to an observed resonance with a linear energy growth for both integer and half integer times the Talbot time. 

Recently, experiments \cite{duffy:qkr} and theory have turned to using atoms originating from a Bose-Einstein Condensate (BEC), which are orders of magnitude colder, less than the heat input of a single photon. Experiments have shown that for these conditions, higher order resonances can be observed \cite{ryu06}, which have been shown to depend sensitively on the initial momentum of the atoms. Furthermore, a range of experiments have exploited ratchets \cite{ratchets,Renzoni06}, and most recently, the dependence of its operation on the initial momentum \cite{sadgrove07,dana08}. 

Here, we present the first measurements of the effect of the initial momentum on the primary quantum resonances in the delta-kicked rotor. We use an ensemble of atoms with a momentum spread much less than a single photon recoil, which is kicked at kick periods corresponding to the quantum resonances. We explore the first three quantum resonances, for two and four kicks. Furthermore, we investigate details of the velocity distribution after a series of kicks.

%\section{Theory}
To describe our system, we are considering a quantum particle of mass $m$ moving in one dimension in a pulsed, spatially periodic potential with period $\lambda/2$. The Hamiltonian is given by
\begin{equation}
 H=\frac{p^2}{2m}+\hbar \phi_d\cos 2k_Lx \sum_{n=1}^N \delta(t-nT)
\end{equation}
where $T$ is the pulse period and $\phi_d$ is the kick strength. It is well known that for a quantum system, strong resonances exist for the Talbot time $T=mT_T$ where the effects of the kicks add constructively, and strong diffraction occurs. Furthermore, at $T=(m+\mbox{$\frac{1}{2}$})\pi/(2\omega_R)$ anti-resonances occur, where the effects of the subsequent kicks add destructively, and only very little diffraction occurs. In the experiment, the delta-function is represented by a block function with a width $\tau\ll T$ and an area $\phi_d$.

We can describe the time evolution of the quantum particle during one kick and the subsequent free evolution period with 
a Floquet operator:
\begin{equation}	
\hat{\mathcal{F}}=\underbrace{e^{-ip^{2}T/{2m\hbar}}}_{=\hat{U}_{\rm free}}
\underbrace{e^{-i\phi_d\cos(2k_Lx)\tau/\hbar}}_{=\hat{U}_{\rm kick}}
\label{eq:timeevolutionofqkickedrotor}
\end{equation}
%\section{Simulation}
We simulate the experiments using the well-known split operator method for the Floquet operator, which is ideally suited to this experiment, with $2^{16}$ grid points. We start the simulation with a minimum uncertainty wave packet, with an initial momentum $\hbar k_i$. The initial wave function is given by
\begin{equation}
\psi(x)=\frac{1}{\sqrt{2\pi}\sigma_w}e^{-\frac{x^2}{2\sigma_w^2}}e^{-ik_ix} 
\end{equation}
The width $\sigma_w=5~\mu$m of the wave packet is chosen so that it covers several wavelengths of the kick laser. The simulation is run with a distribution of initial momenta, corresponding to the initial momentum distribution found in the experiment. These results are added incoherently to obtain a final momentum distribution.

We can condense the effects of the kicks on the velocity to a single number, the variance. This is proportional to the average kinetic energy of the atoms, and can be expressed in recoil energies. The energy is not influenced much by the details of the original momentum distribution, as the typical energies are larger than one recoil, whereas the initial energy is much smaller than one recoil.

For kick periods that are a half integer times the Talbot time, an analytical expression has been derived for the amplitudes of the momentum states with momentum $2j\hbar k$ after $n$ kicks by \cite{Wim03} and \cite{saunders07}:
\begin{equation}
 c_j = J_{j}\left(\phi_d\frac{\sin(n\Upsilon)}{\sin\Upsilon}\right)i^je^{-ij(n+1)\Upsilon}e^{-in\pi\beta^2l} \label{this}
\end{equation}
with $\Upsilon=\frac{1}{2}\pi(1+2\beta)l$, $\beta$ the quasimomentum in units of $2\hbar k_L$, and integer $l=2T/T_T$. 

From equation~\ref{this}, the second order moment of the momentum distribution can be found:
\begin{equation}
 \langle p^2 \rangle = (2 \hbar k_L)^2\sum_{j=-\infty}^\infty \left| c_j \right|^2 (j+\beta)^2 
\end{equation}
The values obtained from equation~\ref{this} are in excellent agreement with those obtained from the simulation, and with the experimental data. 

%\section{Experiment}

In our experiment, we use ultracold atoms sourced from an all-optical BEC of $^{87}$Rb atoms at a temperature of 50~nK, which corresponds to a velocity distribution with a $\sigma\approx0.25$~recoils.  The BEC  is formed in a crossed pair of CO$_2$ laser beams. The $\sim 5\cdot 10^4$ atoms in the BEC are in the $F=1$ hyperfine state, and spread over all magnetic substates. The apparatus is described in detail in \cite{ianpaper}. A pair of counter-propagating diode laser beams, overlapped with the BEC, creates the standing wave potential. In order to reduce mean field effects, the atoms are left free to expand by extinguishing the CO$_2$ laser for 500~$\mu$s after forming the BEC before starting the kick sequence.

The horizontal grating laser beams are derived from a home-built diode laser system. The laser is locked to the $F=2\rightarrow F'$ multiplet in $^{85}$Rb at 780~nm, and thus detuned by 2.2~GHz from the relevant $F=1\rightarrow F=2$ transition in $^{87}$Rb. The laser beam is split 50\%, and both parts are passed through a separate acousto-optic modulator (AOM). The AOMs are driven by an Arbitrary Function Generator (Tektronix AFG-3252), amplified by home-built RF amplifiers. The AOMs are pulsed on simultaneously, but with a tunable frequency difference $\Delta\omega$. This frequency difference gives us an effective initial momentum $p_i$ for the atoms, given by $ p_i/p_{\rm rec}=\Delta \omega/(4\omega_R)$. It should be noted that what is important is the relative phase of the sine waves driving the AOMs in the subsequent pulses, caused by the small frequency difference, not the actual frequency difference.

Care is taken that the laser pulses arrive at the experiment at the same time. A shutter is used in the laser beam to ensure that it is totally extinguished during the evaporation phase to create the BEC. The lasers beams are passed through polarisation preserving single mode fibres to the experiment. The total laser power at the experiment can be varied up to 2~mW in each beam.

The laser beams are focused to a 100~$\mu$m spot at the BEC, in a standing wave configuration. The foci are much larger than the size of the BEC ($\sim10\mu$m), thereby yielding a constant interaction strength over the BEC, whilst increasing the intensity. The increased intensity allows us to perform the experiment at large detunings to suppress spontaneous emission.

An absorption image of the atoms after a 5~ms expansion time is taken to obtain the momentum distributions. The 2D momentum distributions are summed over the width of the cloud to obtain a 1D momentum distribution. In all cases, the experiment is repeated three times. 

%\section{Results and discussion}
%\subsection{two kicks}
\begin{figure}
 \centering
\includegraphics[width=6cm]{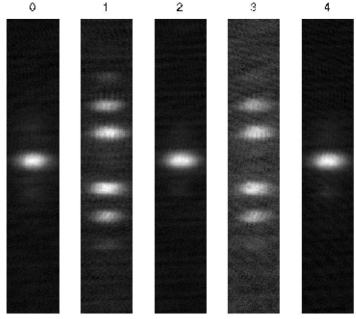}
\caption{Absorption images after two kicks with a kick period of $T=T_T/2$ for different initial momenta in units of photon recoils. \label{fig:absim}}
\end{figure}

In figure~\ref{fig:absim} we show absorption images of the atoms after two kicks at the antiresonance $T=T_T/2$ and after an expansion time of 5~ms for different initial momenta, as set by the frequency difference in the kick beams. At $p=0$ we see almost no diffraction, as expected from the Talbot effect. At $p_i=1~p_{\rm rec}$, as well as at $p_i=3~p_{\rm rec}$, we see strong diffraction, as $T_T/2$ corresponds to a resonance for these atoms. At $p_i=2~p_{\rm rec}$, as well as at $p_i=4~p_{\rm rec}$, there is again no diffraction.

\begin{figure}
 \centering
\includegraphics[width=8cm]{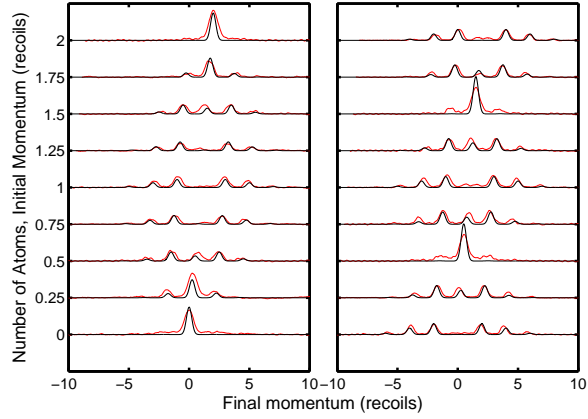}
\caption{The number of atoms as a function of final momentum, for a range of initial momenta from 0 to $2\hbar k$. Two  kicks with period of $T_T/2$  (left) and $T=T_T$ (right) are applied, with experimental profiles (dotted,red)and results from the simulation (full line)  for $\phi_d\approx1$ \label{fig:tr}}
\end{figure}

Next, we vary the initial momentum $p_i$ in 8 steps from 0 to 2~$p_{\rm rec}$. On the left in figure~\ref{fig:tr} we show the resulting 1D momentum distributions, averaged over the three repeats of the experiment. It can be seen that at zero velocity, there is indeed an anti-resonance and very little diffraction occurs. However, it turns into a resonance at $p_i= p_{\rm rec}$, with significant diffraction, and back to an anti-resonance at $p_i=2p_{\rm rec}$ with little diffraction. Also shown in the figure are the results of the simulation. When experiments are performed with a broad initial momentum distribution, an average over many different $v_i$ yields an observed resonance at $T=T_T/2$, as has been shown in many publications, see, e.g., reference \cite{oskay00}. 

On the right in figure~\ref{fig:tr} we show the resulting 1D momentum distributions for two kicks at a spacing of $T=T_T$, for varying $p_i$, both experiment and simulation.
At $p_i=0$, there is a resonance as expected, with significant diffraction. However, at $p_i=0.5 p_{\rm rec}$, this turns into an anti-resonance, back to a resonance at $p_i=p_{\rm rec}$, back to an antiresonance at $p_i=1.5 p_{\rm rec}$ and finally to a resonance at $p_i=2 p_{\rm rec}$. Again, averaging over a range of initial velocities will show the same resonance at $T=T_T$, even though the dependence on the initial velocity cycles at twice the rate. At $T=3T_T/2$, the cycle of the amount of diffraction varying with the initial velocity is at three times the rate we see for $T=T_T/2$, and so on. The simulated curves show good agreement in terms of the relative heights of all the diffraction peaks.
 \begin{figure}
 \centering
 \includegraphics[width=9cm]{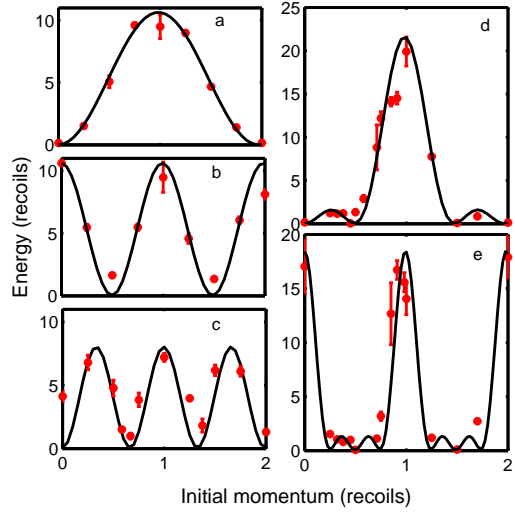}
 \caption{On the left, the final energy as a function of the initial momentum for two kicks at a pulse period of 33.15 $\mu$s (a), 66.3~$\mu$s (b) and 99.45~$\mu$s (c). On the right, the energy after four kicks at a period of $T_T/2$ (d) and $T_T$ (e). The experimental data and uncertainties (markers) are shown as well as the results from the simulation (line). The kick strength $\phi_d\approx1$ for these figures, except that $\phi_d=0.9$ for (c)}
 \label{fig:sigs2}
\end{figure}

A characteristic feature of these velocity distributions is the variance of the momentum distribution, or the mean kinetic energy in units of the recoil energy $E_r=\hbar\omega_R$. In figure~\ref{fig:sigs2} we show the variation of the energy on the initial momentum for $T=T_T/2$ (a), $T=T_T$ (b) and $T=3T_T/2$ (c), experiment and simulation. For $T=3T_T/2$, the energy varies at three times the rate for $T=T_T/2$. For the data points, the averaged profile is fitted to a number of Gaussians, one for each diffraction order, to obtain the energies. The uncertainties are generated by fitting each experiment to obtain the energy and taking the standard deviation. It was found that a direct numerical variance of the velocity distributions gave similar results, but much less consistent due to its sensitivity to small noise peaks at large momenta, and consequently these values were not used.

For a larger number of kicks, it is well known that the width of the quantum resonance, when plotted versus the kick period, gets smaller as the number of kicks gets larger. Expecting a similar behavior, we investigate the dependence of the energy on the initial velocity for four kicks. In figure~\ref{fig:sigs2} we show the energy as a function of the initial velocity for four kicks, for both the anti-resonance (d) and the resonance (e). For $T=T_T/2$ (d) there is an anti-resonance at zero momentum, yielding small energies. There is a small maximum at an initial momentum of $\frac{1}{4}$ recoils, decreasing again to close to zero at $\frac{1}{2}$ recoils. At an initial momentum of one recoil, there is a stronger maximum in the energy, decaying again to small energies at $\frac{3}{2}$ before another smaller maximum and returning to small energies at two recoils initial momentum. All these features, prominent in our calculations, are reproduced by the experiment.  

For $T=T_T$ (e) a larger number of oscillations is observed in the simulated curve, along with more narrow maxima. The experiment reproduces the maxima, but at current cannot resolve the small-period oscillations for these settings.

\begin{figure}
 \centering
\includegraphics[width=8cm]{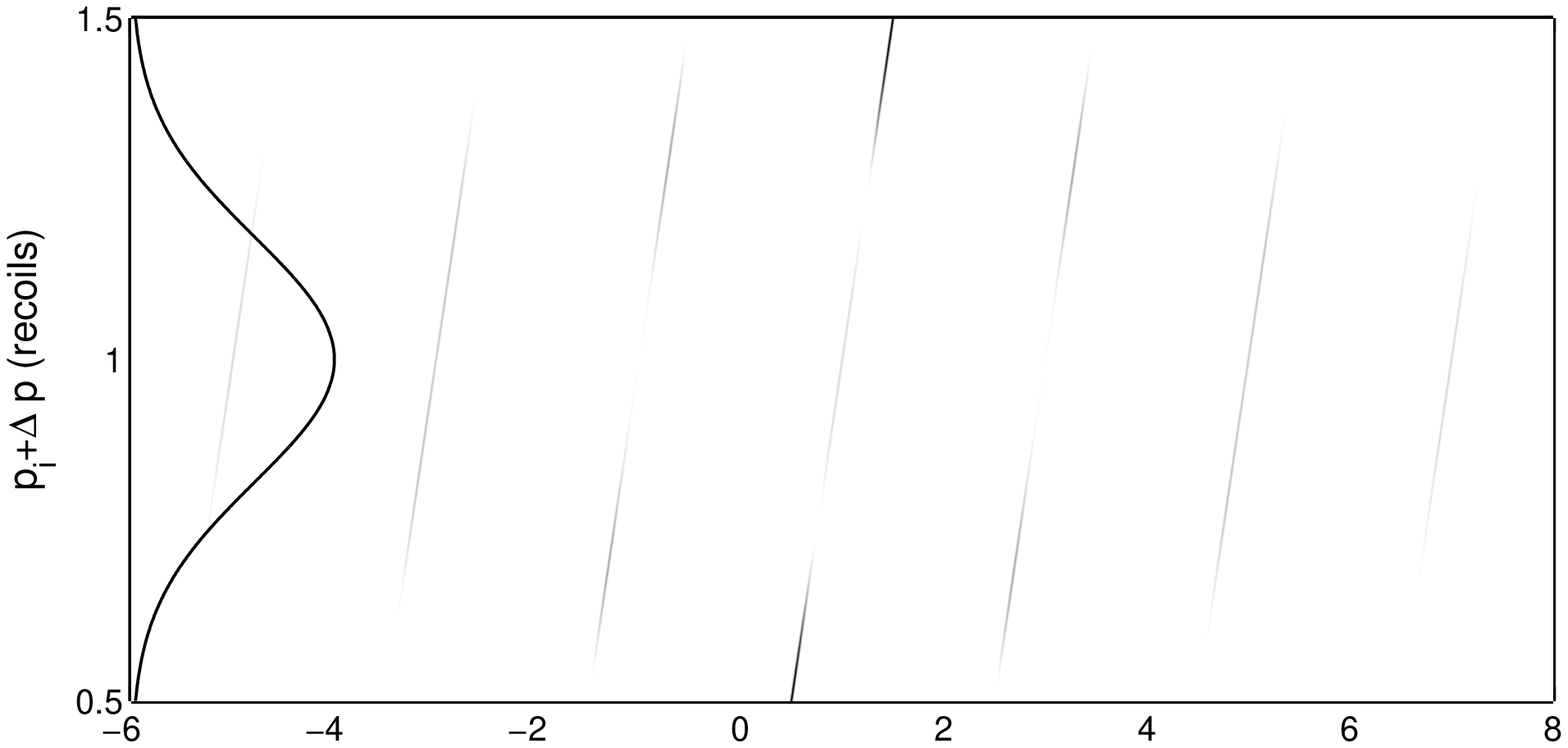}
\includegraphics[width=8cm]{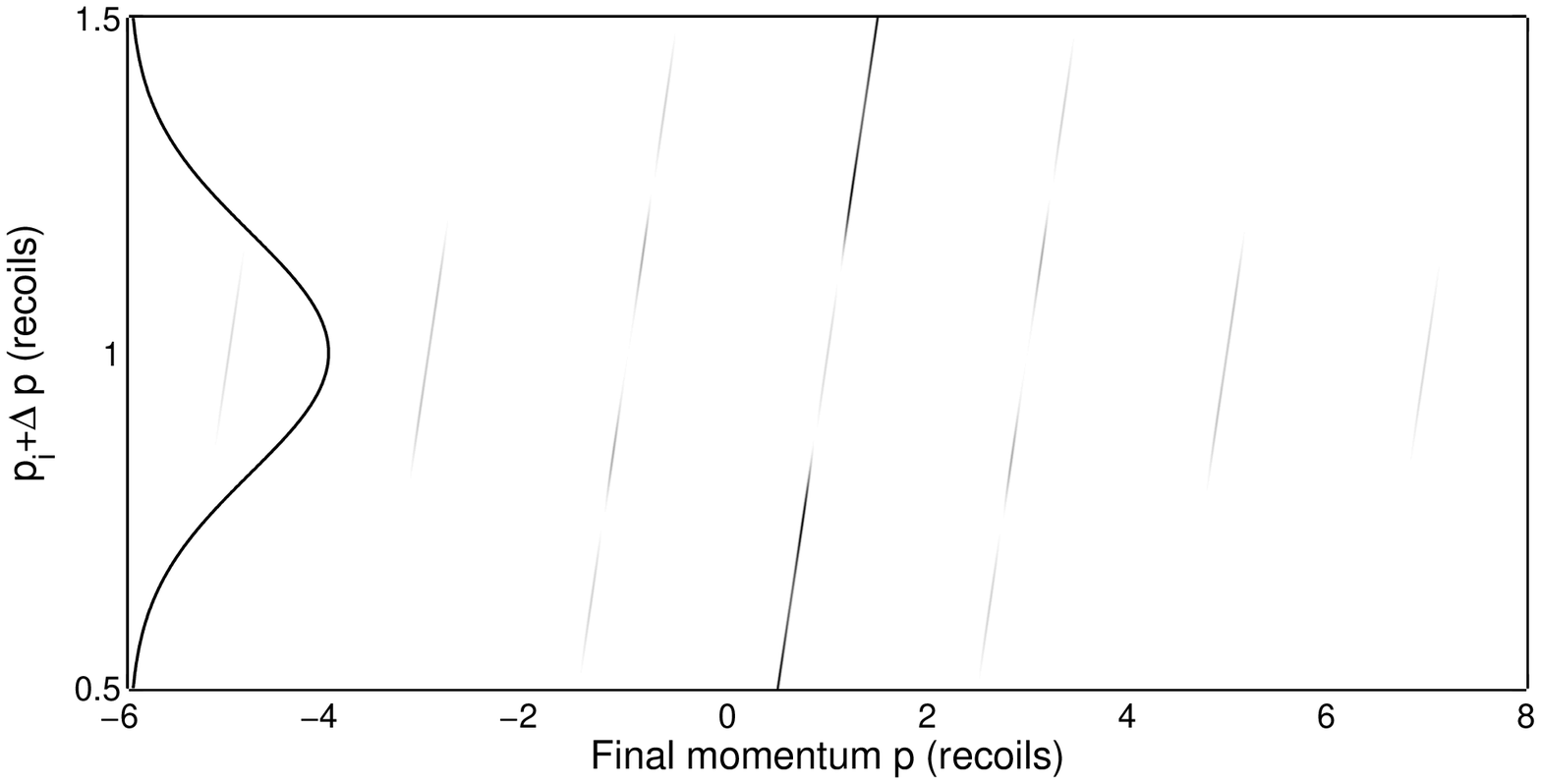}
\caption{The probability of finding atoms (grayscale) at a final momentum (horizontal axis) for a range of initial momenta (vertical axis). Also shown is a Gaussian momentum distribution which represents the distribution of the initial BEC. This is for four kicks at $T=T_T/2$ (top) and $T=T_T$(bottom),  average initial momentum $\langle p_i\rangle=p_{\rm rec}$ and $\phi_d=1$ \label{fig:theory}}
\end{figure}

Now, we will take a closer look at the momentum distribution after four kicks, with an initial momentum of one recoil. The initial momentum distribution from the BEC has a certain width $\sigma=0.18$ recoils, which qualitatively alters the momentum distribution after the kick sequence. Ideally, the initial momentum distribution is a delta function at $p_i$. In reality, the initial momentum distribution has a width, and there are also some atoms at a momentum $p_i+\Delta p$, where $\Delta p$ is much smaller than one recoil. 
To illustrate this, we performed a simulation for a range of initial momenta.
In figure~\ref{fig:theory} we show the probability (grayscale) of finding the atom at a final momentum (horizontal axis) for a range of initial momenta for both the antiresonance and the resonance. We also display the initial momentum distribution of the BEC (black line).
The sensitivity of the system to the initial momentum translates into peaks coming into existence at a small offset momentum $p_i+\Delta p$ that are not there at $p_i$ ($p_i$ is equal to one recoil in this case), or, conversely, a peak disappearing at $p_i+\Delta p$ which was present at $p_i$. 
 
\begin{figure}
 \centering
\includegraphics[width=8cm]{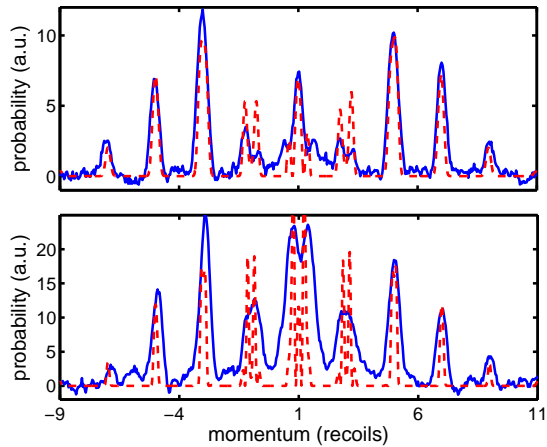}
\caption{The momentum distributions after four kicks for $T=T_T/2$ (top) and $T=T_T$ (bottom) with the experimental profiles (full lines) and the simulation (dashed lines). In both cases, $\langle p_i\rangle=p_{\rm rec}$. Again, $\phi_d\approx 1$ \label{fig:dist4}} 
\end{figure}

 As mentioned earlier, we model the effect of a finite width of the initial momentum distribution by carrying out the simulation for a range of $\Delta p$, and we subsequently add the momentum distributions thus obtained incoherently, weighted by the height of the momentum distribution of the original BEC at $\Delta p$. In effect, to obtain the final momentum distribution, we take horizontal profiles in figure~\ref{fig:theory}, and sum these, weighed by the initial momentum distribution.

These momentum distributions are shown in figure~\ref{fig:dist4}. The difference in width between the different ``diffraction orders'' can easily be observed and related to the separation of one velocity peak into multiple peaks in the simulation. For instance, the peak at a momentum $p=1$ recoil is split into three, with varying heights in the two situations. This would be caused by the peak disappearing for $\Delta p=0.2$ recoils, and subsequently re-appearing for $\Delta p=0.4$. The latter peaks appear in the final momentum distribution even though they are not strongly weighted by the initial velocity distribution. 

The peaks at $p=(3,-1)$ recoils are split into two, which would be caused by peaks appearing at a small $\Delta p$. On the other hand, the peaks at $p=-(5,7)$ recoils are considerably narrower than the initial momentum distribution, which is caused by the fact that this diffraction peak only appears for very small $\Delta p$. These observations can be verified in figure~\ref{fig:theory}, by drawing a horizontal line at fixed $\Delta p$ and finding the diffraction orders.

At a larger number of kicks, the simulations show that the width of resonances as a function of initial momentum will get smaller. Determining the energy after a number of kicks could be used to measure the velocity distributions of an atomic sample with great accuracy. As it is influenced by the velocity only, it is not influenced by the initial cloud size as is the time-of-flight method.

%\section{Conclusions}
In summary, we have for the first time experimentally demonstrated the dependence of the quantum resonances in the delta-kicked rotor on the initial velocity of the atoms. We show a sinusoidal dependence of the energy on the initial momentum for two kicks, and a more complex behavior with the same period at four kicks. We propose to use the momentum dependence of the kicked rotor to select narrow parts of an initial momentum distribution, or to measure a given momentum distribution with great accuracy. 

The authors would like to acknowledge fruitful discussions with Scott Parkins regarding this work.
%\section*{References}
%\bibliographystyle{epl}
%\bibliographystyle{apsrev}
%\bibliography{kickedrotors}

\end{document}